\documentclass[conference]{IEEEtran}

\ifCLASSINFOpdf
   \usepackage[pdftex]{graphicx}
   \graphicspath{{../pdf/}{../jpeg/}}
   \DeclareGraphicsExtensions{.pdf,.jpeg,.png}
\else
   \usepackage[dvips]{graphicx}
   \graphicspath{{../eps/}}
   \DeclareGraphicsExtensions{.eps}
\fi

\usepackage{array}
\usepackage{url}
\usepackage{float}

\begin{document}

        \title{Overlapping Community Structure in Co-authorship Networks: a Case Study}

        \author{\IEEEauthorblockN{Malek Jebabli}
        \IEEEauthorblockA{Universit{\'e} de Bourgogne\\
        Dijon, France\\
        Email: Malek.Jebabli\\@u-bourgogne.fr}
        \and
        \IEEEauthorblockN{Hocine Cherifi}
        \IEEEauthorblockA{Universit{\'e} de Bourgogne\\Dijon, France\\
        Email: hocine.cherifi\\@u-bourgogne.fr }
        \and
        \IEEEauthorblockN{Chantal Cherifi}
        \IEEEauthorblockA{Universit{\'e} Lumi{\'e}re\\Lyon, France \\
        Email: Chantal.BonnerCherifi\\@univ-lyon2.fr }
        \and
        \IEEEauthorblockN{Atef Hamouda}
        \IEEEauthorblockA{University Tunis El-Manar\\
        Tunis, Tunisia\\
        Email: atef\_hammouda\\@yahoo.fr }}

\maketitle

\begin{abstract}
        Community structure is one of the key properties of real-world complex networks. It plays a crucial role in their behaviors and topology. While an important work has been done on the issue of community detection, very little attention has been devoted to the analysis of the community structure. In this paper, we present an extensive investigation of the overlapping community network deduced from a large-scale co-authorship network. The nodes of the overlapping community network represent the functional communities of the co-authorship network, and the links account for the fact that communities share some nodes in the co-authorship network. The comparative evaluation of the topological properties of these two networks shows that they share similar topological properties. These results are very interesting. Indeed, the network of communities seems to be a good representative of the original co-authorship network. With its smaller size, it may be more practical in order to realize various analyses that cannot be performed easily in large-scale real-world networks.
\end{abstract}
\IEEEpeerreviewmaketitle
\section{Introduction}
        Many real-world systems take the form of complex networks, i.e. a large set of items or nodes, with a set of links or edges between them. These networks, represented by graphs, share some common topological properties. Indeed, most real-world complex networks are scale-free and exhibit the ''small-world'' property. Furthermore, they are characterized by a high clustering coefficient and a well-defined community structure. Uncovering the community structure is crucial to the understanding of the structural and functional properties of real-world complex networks. As there is no universal definition of a community, many algorithms have been proposed \cite{Fortunato2010a}, \cite{Xie2011}, \cite{hocine2014}. They usually rely on the fact that nodes in the same community are more densely connected to each other than to the rest of the network. We can distinguish two types of community detection algorithms. Non-overlapping community detection deals with the case where nodes belong to only one community while in overlapping community detection, nodes can belong to multiple communities. Most real networks and particularly social networks are well defined by overlapping and nested communities. Indeed, each of us belongs to numerous communities related to our multiple activities. Based on this assumption, we can define a network of communities where the nodes are the communities and the links represent the interactions between the communities. While there has been numerous works on the community detection issue, few authors tackled the analysis of the overlapping community structure. In this context, the contribution of Palla et al \cite{Palla2005} is one of the most influential work. In this work, they introduced four basic quantities, namely the membership number, the overlap size, the community size and the community degree in order to characterize the global organization of networks with overlapping communities. The membership number is the number of communities to which a node belongs. The overlap size measures the number of nodes shared by any two communities. The community size is the number of node in a community, and the degree of a community is the number of communities overlapping with it. The authors studied a co-authorship network, a network of word associations related to cognitive sciences and a molecular-biological network of protein-protein interactions. In this work, communities are identified using a clique percolation community detection algorithm. The authors found that the community degree distribution exhibits an exponential decay followed by a Power-law tail. The three other distributions are well approximated by a Power-law. In addition, they analyzed the networks of communities (where the nodes are the communities, and there is a link between two communities if their share at least one node). They showed that these networks are characterized by a high clustering coefficient. Note that the degree of a community corresponds to the degree of a node in the overlapping community networks. There are some other interesting studies \cite{Lancichinetti2010}, \cite{Ahn2010}, \cite{Tibely2011} on overlapping community structure analysis, however, they suffer from two main drawbacks. First of all, they rely on a community detection algorithm to identify the community structure. As there is no universal method, results are very sensitive to this issue. Furthermore, the size of the networks used in these studies are relatively small. Our work responds to these criticisms. In order to get rid of the community detection issue, we use a network with ground truth community structure. In addition, we investigate a large-scale network with hundreds of thousands of nodes and links. To our knowledge, large-scale networks with ground truth community structure have not previously been used in order to analyze the overlapping community networks. Our aim is to investigate the relationship between the original co-authorship network and its overlapping network of communities. The rest of the paper is organized as follows. Section 2 gives the background about the overlapping community graph and the properties under investigation. Section 3 presents the results of our analysis. Finally, in Section 4 we present conclusions.

\section{Topological Characteristics}
        Structural analysis of complex networks can be performed at different scales ranging from the microscopic to the macroscopic level. In this study, we focus on macroscopic level. In the macroscopic level, statistical measures are used to summarize some of the overall network features. In this section, we briefly recall some measures commonly used to capture, in quantitative terms, the networks organizing principles. A complex network is modeled by a graph where the nodes are individuals connected by links, which mimic their interactions. Let $G=(V,E)$ represents a connected, undirected, and unweighted graph where $V$ is the set of $n$ nodes and $E$ is the set of $m$ links of $G$. For the macroscopic topological properties, the \emph{small-world} property refers to the low average distance value between any two nodes of a network. The \emph{global clustering coefficient} reflects the tendency of link formation between neighboring nodes in a network. \emph{Degree distribution} measures the statistical repartition of the network nodes’ degrees. For a large number of networks, it can be adequately described by a Power-law distribution $(P(k)\sim k^{-\alpha})$, where $\alpha$ is a positive exponent. These networks are often referred as ''scale-free networks'' because their degree distribution does not depend on their size. Related experimental studies show that the exponent value of the Power-law usually ranges from two to three. The \emph{degree correlation} measures the tendency of nodes to associate with other nodes sharing the same characteristics and especially the same degree values. In assortative networks, the nodes tend to associate with their connectivity peers, and the degree correlation is positive. In disassortative networks, high-degree nodes tend to associate with low-degree ones, and the degree correlation is negative. Social networks appear to be assortative while information, technological and biological networks appear to be disassortative. \emph{The average clustering coefficient as a function of node degree} gives details of a network’s triangular clustering structure. For a large number of networks, it can be represented by a Power-law distribution \cite{Cheng2009}. The \emph{hop-distance} for the network is the set of pairs $(d,g(d))$ where for each natural number  $d$, $g(d)$   denote the fraction of connected node pairs whose shortest connecting path has length at most $d$. The hop plot represents the distribution of pairwise distances in a network.

\section{Experimental Results}
\subsection{The data and the overlapping community network basic structure}
        For our experiments, we use the Digital Bibliography and Library Project (DBLP) scientific collaboration network from the Stanford Large Network Dataset Collection (SNAP). This is an undirected, unweighted network where nodes represent authors and links connect nodes that have co-authored at least one publication. Publications are used in order to define the ground-truth communities. In other words, authors who publish in the same journal or conference belongs to the same community. As an author usually publishes in different media, he belongs to several communities. This network is complete, therefore there is no bias on its statistics introduced by any sampling procedure. It contains $317080$ nodes and $1049866$ links that are distributed into $13477$ communities. This network is made of one large connected component (the small components have been removed). The overlapping community network is made of a giant component, few small components and isolated nodes. The great majority of the nodes belong to the giant component $(99.43\%)$. Isolated nodes represent only $0.12\%$ while small components account for $0.45\%$. In the following, we restrict our attention to the giant component of the overlapping community network. We refer to it with the abbreviation DBLP*, while DBLP represents the giant component of the co-authorship network. For short, we use the word "network" when referring to the giant components. DBLP* contains $13151$ nodes and $598872$ links. The proportion of links of this component is equal to $99\%$ of the totality of links of the overlapping community network. It is denser that the co-authorship network. Indeed, its density is equal to $0.018$ as compared to $1.04*10^{-5}$  for DBLP. This is due to the high number of overlapping between the communities.

        \newcommand{\specialcell}[2][c]{%
          \begin{tabular}[#1]{@{}c@{}}#2\end{tabular}}

        \subsection{Degree distribution}
\begin{figure}[!htbp]
        \centering
        \includegraphics[width=2.5in]{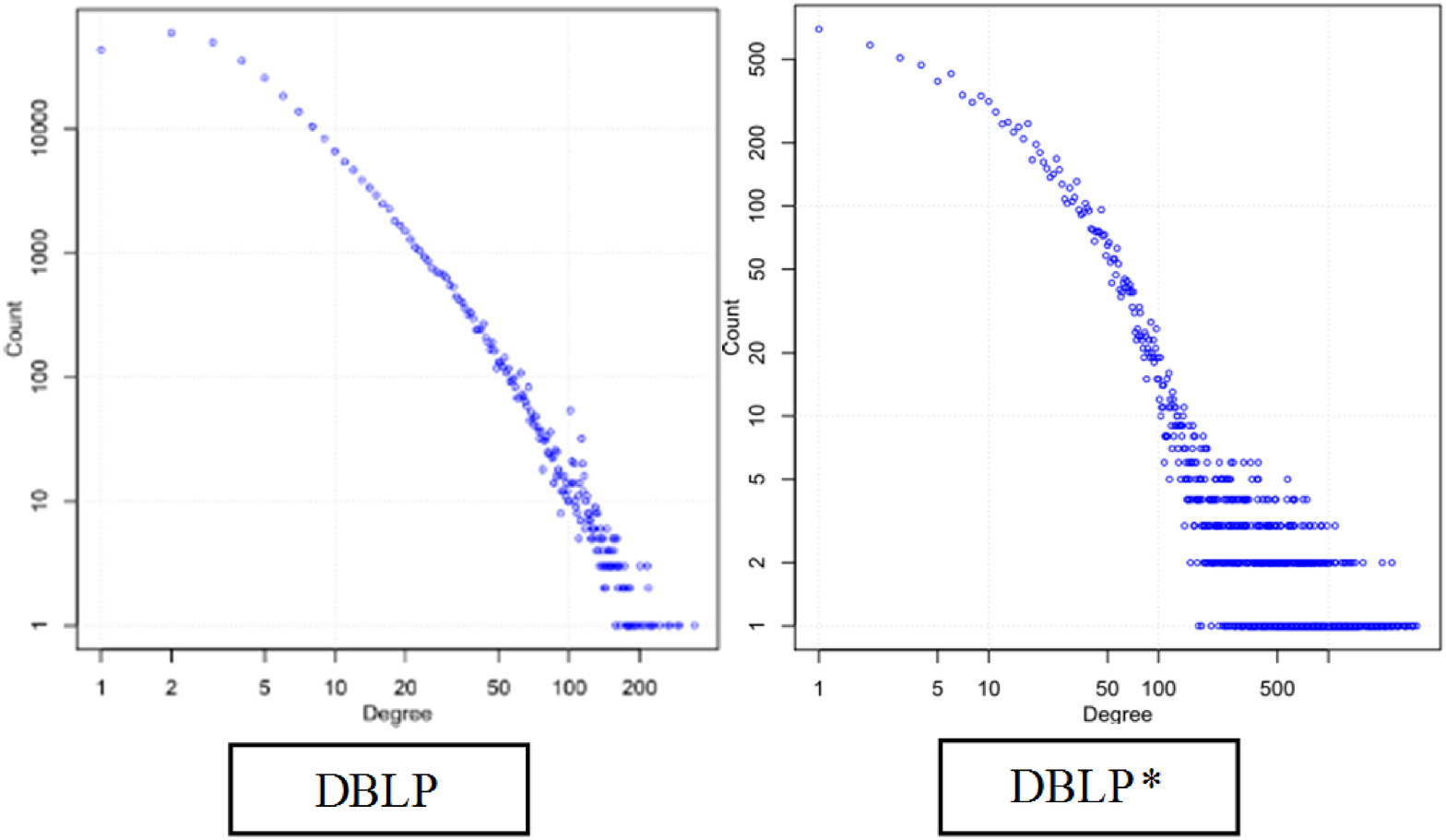}
        \caption{Log-log degree distribution of the co-authorship network DBLP and the associated networks of communities DBLP*}
        \label{fig1}
        \end{figure}

        \figurename \ref{fig1} reports the empirical degree distribution of the original network together with the overlapping community network. It appears that the shape of the degree distribution of the original network and the overlapping community networks are very similar. The Kolmogorov–Smirnov test has been used to fit different probability distribution (Power-law, Beta, Cauchy, Exponential, Gamma, Logistic, Log-normal, Normal, Uniform, and Weibull). The results reported in Table \ref{table1} demonstrates that the power-law is the best fit. The estimates of the Power-law exponent is $\alpha=3.2$ for DBLP and $\alpha=2.98$ for DBLP*. These results are consistent with the values usually observed.
        However, we note a significant increase of the degree range for DBLP*. Indeed, the maximum degree value is equal to $3300$ for DBLP* as compared to $343$ for DBLP. Furthermore the mean degree value of DBLP* is equal to $90.3$ while it is equal to $6.6$ for DBLP. To summarize, the degree distribution of both networks follow a power law with an increase in the overall order of the degree from the real networks to the networks of communities.

        \begin{table}[!htbp]
        \renewcommand{\arraystretch}{1.3}
        \caption{KS-test values for the Degree distribution. The distribution under test are the Power-law (pl), Beta (bet), Cauchy (cau), Exponential (e), Gamma (gm), Logistic(log), Log-normal (ln), Normal (n), Uniform (u), and Weibull (wb)}
        \label{table1}
        \centering
        \begin{tabular}{|p{.55cm}|*{10}{|p{.26cm}|}}
        \cline{2-11}
        \multicolumn{1}{c|}{} &  PL & BET & CAU & E & GM & LOG & LN & N & U & WB \\
        \hline
         DBLP&0.03&0.29&0.22&0.14&0.16&0.24&0.1&0.29&0.89&0.15\\
        \hline
        DBLP*&0.03&0.22&0.26&0.32&0.17&0.35&0.04&0.36&0.85&0.1\\
        \hline
        \end{tabular}
        \end{table}

        \subsection{The average clustering coefficient as a function of nodes degree}
\begin{figure}[!htbp]
        \centering
        \includegraphics[width=2.5in]{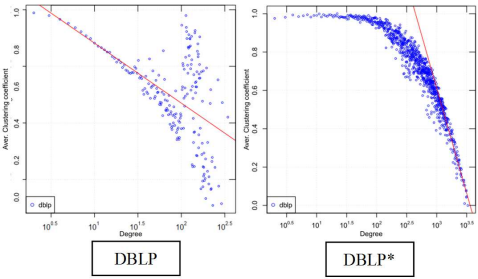}
        \caption{Average clustering coefficient distribution as a function of degree for the co-authorship network DBLP and the associated networks of communities DBLP*}
        \label{fig2}
        \end{figure}

        For each degree $k$, we calculate the average clustering coefficient. Then we sort these values as a function of the degree, so we have the average clustering coefficient as a function of nodes degree (see \figurename \ref{fig2}). For the DBLP network, the beginning of the distribution behaves as a Power-law distribution. However this is not the case for the the tail. Nevertheless, according to the KS-test results reported in Table \ref{table2}, the best fit is the power law with an exponent value $\alpha=3.06$. It is followed by the Weibull distribution with the shape parameter value $k=1.14$ and the scale parameter value $\lambda=0.2$ . For DBLP*, the results are not very conclusive. The best fit is the Weibull distribution with the shape parameter value $k=2.04$ and the scale parameter value $\lambda=0.33$. The log-normal and the power law distribution share the second position with a KS value of $0.14$ as compared to $0.1$ for the weibull distribution (see Table \ref{table2}). When we look at the shape of the distributions reported in \figurename \ref{fig2}, we can observe that they are quite different. The power law seems to be a good fit for low degree values for DBLP while it is true for high degree values for DBLP*.

        \begin{table}[!htbp]
        \renewcommand{\arraystretch}{1.3}
        \caption{KS-test value for The average clustering coefficient as a function of nodes degree. The distribution under test are the Power-law (pl), Beta (bet), Cauchy (cau), Exponential (e), Gamma (gm), Logistic(log), Log-normal (ln), Normal (n), Uniform (u), and Weibull (wb)}
        \label{table2}
        \centering
        \begin{tabular}{|p{.55cm}|*{10}{|p{.26cm}|}}
        \cline{2-11}
        \multicolumn{1}{c|}{} &  PL & BET & CAU & E & GM & LOG & LN & N & U & WB \\
        \hline
        DBLP&0.04&0.12&0.13&0.2&0.14&0.11&0.22&0.1&0.23&0.06\\
        \hline
        DBLP*&0.14&0.15&0.19&0.22&0.15&0.14&0.34&0.17&0.42&0.1\\
        \hline
        \end{tabular}
        \end{table}

\subsection{Hop Distance distribution}

\begin{figure}[!htbp]
        \centering
        \includegraphics[width=2.5in]{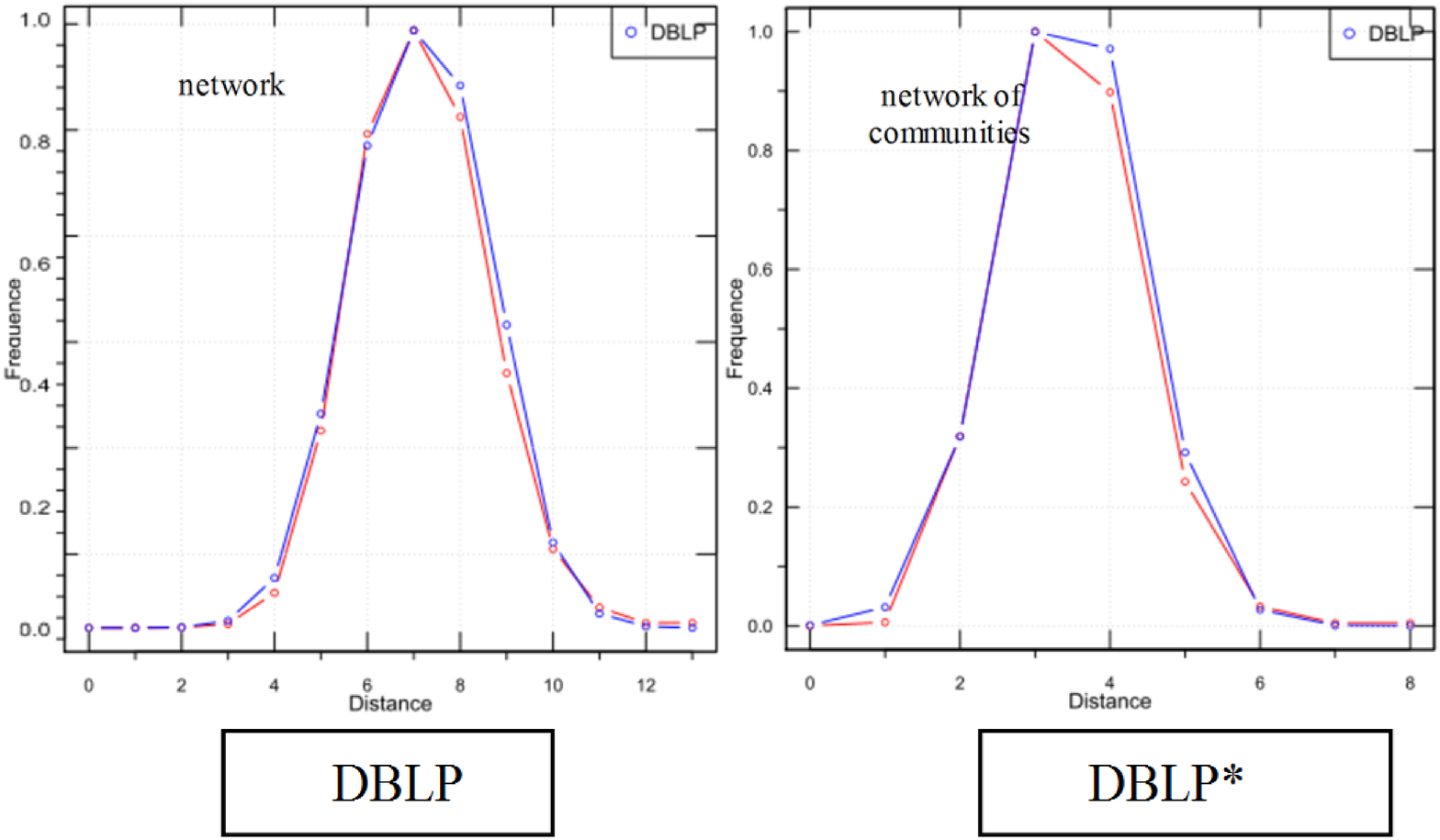}
        \caption{Hop distance distribution for the co-authorship network DBLP and the associated networks of communities DBLP*}
        \label{fig3}
        \end{figure}

        The distributions of pairwise distances reported in \figurename \ref{fig3} are very similar. According to the KS test, in Table \ref{table3}, the best fit is the normal distribution with a mean value $\mu =7$ and standard deviation $\sigma = 0.08$ for DBLP as compared to $\mu =2.93$ and standard deviation $\sigma = 0.07$ for DBLP*. Consequently, except for the shift of the mean values between the two distributions we can conclude that both networks exhibit the same behavior.

        \begin{table}[!htbp]
        \renewcommand{\arraystretch}{1.3}
        \caption{KS-test value for Hop Distance distribution. The  distribution under test are the Power-law (pl), Beta (bet), Cauchy (cau), Exponential (e), Gamma (gm), Logistic(log), Log-normal (ln), Normal (n), Uniform (u), and Weibull (wb)}
        \label{table3}
        \centering
        \begin{tabular}{|p{.55cm}|*{10}{|p{.26cm}|}}
        \cline{2-11}
        \multicolumn{1}{c|}{} &  PL & BET & CAU & E & GM & LOG & LN & N & U & WB \\
        \hline
        DBLP&0.42&0.17&0.13&0.7&0.74&0.26&0.11&0.05&0.24&0.28\\
        \hline
        DBLP*&0.34&0.2&0.16&0.64&0.88&0.14&0.08&0.03&0.47&0.2\\
        \hline
        \end{tabular}
        \end{table}

\subsection{Miscellaneous properties }

        In Table \ref{table4}, the average shortest path, together with the diameter, the global clustering coefficient and the degree correlation of DBLP and DBLP* are reported. Both networks exhibit small-world characteristics.  With an average shortest path around $2.5$, most nodes are just a few edges away on average. Note that their values differ by less than $7\%$. The diameter, which is the longest of all the shortest paths in a network, is identical for both networks. This strict equality is probably a coincidence, nevertheless, this observation confirms that the properties of the two networks are very similar. When we consider the average clustering coefficient, the values reported differ by less than $5\%$. Note that, according to da Fortuna Costa et al \cite{Costa2011},  networks with a transitivity above $0.3$ are considered highly transitive. So we can conclude that DBLP and DBLP* are realistic in terms of transitivity. Considering the degree correlation, the networks show different behavior. The absolute values of the degree correlation are very close, but the sign of the correlations are different. It is positive for DBLP and negative for DBLP*. Positive values indicate relationships between nodes of similar degree, while negative values indicate relationships between nodes of different degree. In other words, like many social networks DBLP is assortative while DBLP* appear to be disasortative, as technological and biological networks. Indeed, authors tend to get acquainted to their pairs while communities have a tendency to evolve towards their maximum entropy state which is usually disassortative.

        \begin{table}[!htbp]
        \renewcommand{\arraystretch}{1.3}
        \caption{Global properties of DBLP and DBLP*}
        \label{table4}
        \centering
        \begin{tabular}{|c||c||c||c||c|}
        \hline
         &  \specialcell{Average \\ shortest path} & Diameter & \specialcell{Average clustering \\ coefficient} & \specialcell{Degree \\ correlation} \\
        \hline
         DBLP&2.76&21&0.39&0.26\\
        \hline
        DBLP*&2.58&21&0.41&-0.28\\
        \hline
        \end{tabular}
        \end{table}

\begin{figure}[!htbp]
        \centering
        \includegraphics[width=1.5in]{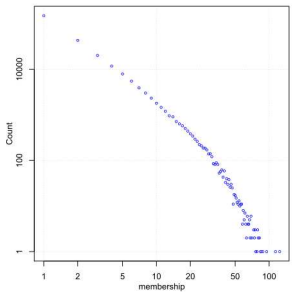}
        \caption{Log-log membership distribution of DBLP}
        \label{fig4}
        \end{figure}

\begin{figure}[!htbp]
        \centering
        \includegraphics[width=1.5in]{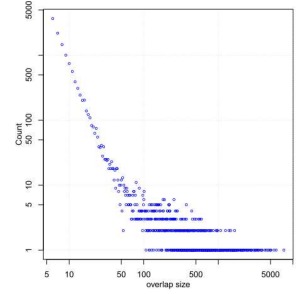}
        \caption{Log-log overlaps size distribution of DBLP}
        \label{fig5}
        \end{figure}

\begin{figure}[!htbp]
        \centering
        \includegraphics[width=1.5in]{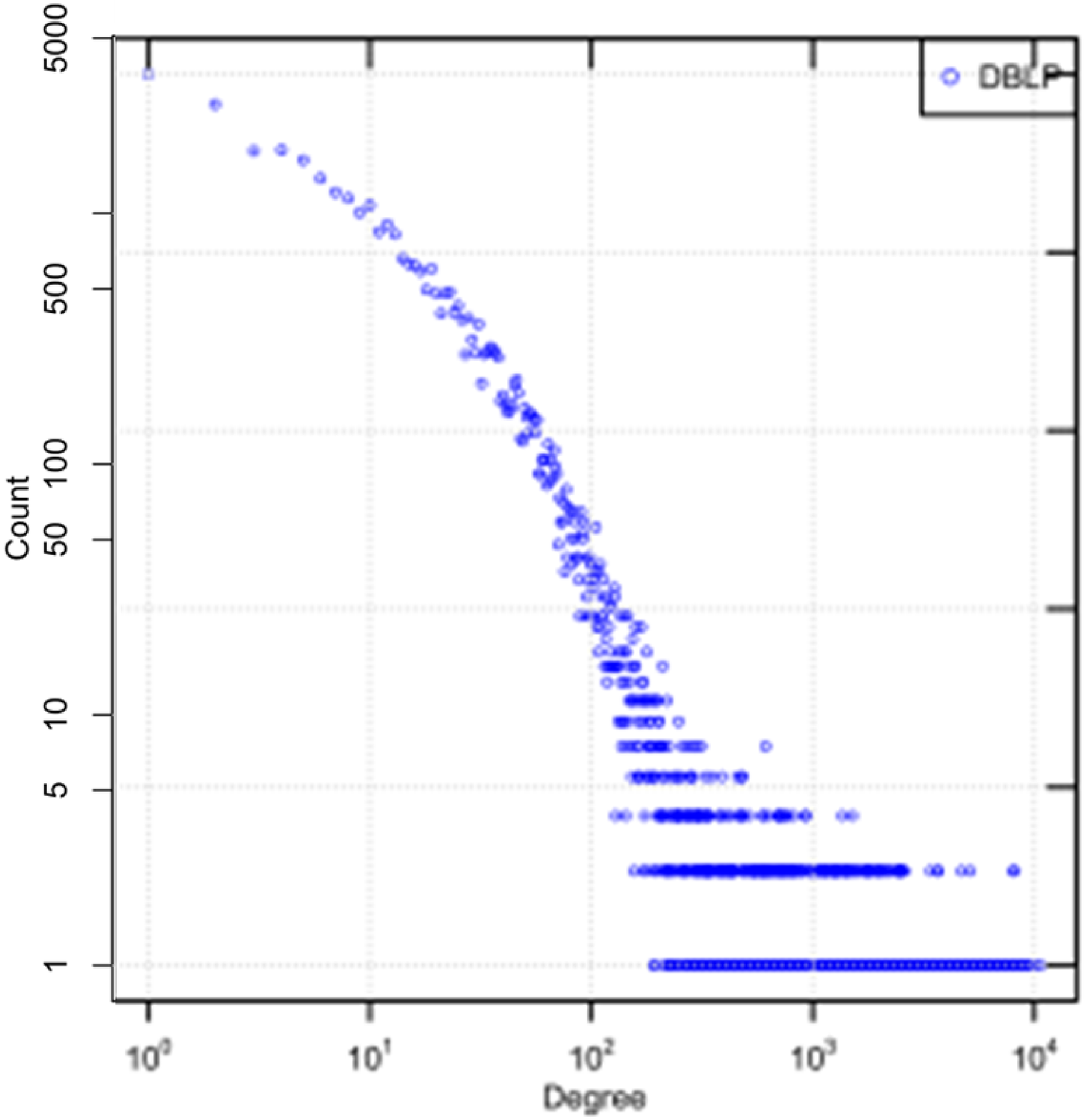}
        \caption{Log-log community degree  distribution of DBLP}
        \label{fig6}
        \end{figure}

        Note that we have computed the distribution of the measures introduced by Palla (membership number of a node, overlap size between communities, and  community size) for DBLP (see \figurename \ref{fig4},\ref{fig5},\ref{fig6}). The best fits for all these properties is the power law (see Table \ref{table5}). These results are in accordance with previous experiments.

        \begin{table}[!htbp]
        \renewcommand{\arraystretch}{1.3}
        \caption{KS-test value for membership, overlap size and  community size  of DBLP. The distribution tested  are the Power-law (pl), Beta (bet), Cauchy (cau), Exponential (e), Gamma (gm), Logistic(log), Log-normal (ln), Normal (n), Uniform (u), and Weibull (wb)}
        \label{table5}
        \centering
        \begin{tabular}{|p{0.7cm}|*{10}{|p{.25cm}|}}
        \cline{2-11}
        \multicolumn{1}{c|}{} &  PL & BET & CAU & E & GM & LOG & LN & N & U & WB \\
        \hline
        \specialcell{Member-\\ship}&0.03&0.41&0.17&0.3&0.32&0.34&0.33&0.34&0.89&0.31\\
        \hline
        \specialcell{Overlap\\size}&0.06&0.35&0.19&0.56&0.37&0.43&0.27&0.43&0.93&0.36\\
        \hline
        \specialcell{Commu.\\size}&0.04&0.29&0.14&0.66&0.35&0.36&0.47&0.31&0.89&0.77\\
        \hline
        \end{tabular}
        \end{table}

\section{Conclusion}

        We have performed a detailed analysis of a co-authorship network with functionally defined communities. Focusing on the overlapping network of community, we show that there is a relative similarity between the original co-authorship network and the overlapping network of communities. Indeed, the degree distribution of both networks follows a Power-law. Additionally, the hop distance is well approximated by a Gaussian distribution in both cases. Finally, the average shortest path, the diameter and the global clustering coefficient are very similar. Differences between the two networks appear when we consider the average clustering coefficient as a function of nodes degree and the degree correlation. Indeed the co-authorship network is assortative while the overlapping community network is disassortative. Furthermore, the distribution of the average clustering coefficient as a function of nodes degree exhibit different behavior. The similarities between the networks reported in this paper are very important. Indeed, manipulating the network of communities is a very simple task as compared to the original network. Furthermore, the network of communities can be used to derive a quality measure in order to evaluate the various community detection algorithms. These preliminary results are promising, therefore we intend to extend this work in two directions. First of all we want to refine the analysis by considering additional properties such as centralities. Secondly, we plan to investigate alternative networks in order to better understand the relationship between the original network topology and the community structuration.

 \bibliographystyle{IEEEtran}
\bibliography{IEEEabrv,./biblio}
\end{document}